# Angle-resolved photoemission spectroscopy and its applications to topological materials

B. Q. Lv,[1,2]  T. Qian,[1,3,4*]  and H. Ding[1,3,4#]

[1] *Beijing National Laboratory for Condensed Matter Physics and Institute of Physics, Chinese Academy of Sciences, Beijing 100190, China*

[2] *Department of Physics, Massachusetts Institute of Technology, Cambridge, MA 02139, USA.*

[3] *CAS Center for Excellence in Topological Quantum Computation, University of Chinese Academy of Sciences, Beijing 100190, China*

[4] *Songshan Lake Materials Laboratory, Dongguan, Guangdong 523808, China*

[#] dingh@iphy.ac.cn,  [*] tqian@iphy.ac.cn

**Angle-resolved photoemission spectroscopy (ARPES), an experimental technique based on the photoelectric effect, is arguably the most powerful method for probing the electronic structures of solids. The last decade has witnessed significant progresses in ARPES: rapid developments of soft X-ray ARPES, ultrafast time-resolved ARPES, spin-resolved ARPES, and spatially resolved ARPES, as well as significant improvements in energy and momentum resolutions. Consequently, ARPES has emerged as an indispensable experimental probe in studying topological materials, which harbor characteristic nontrivial electronic structures in the bulk and on the boundary. Over the past few years, ARPES has played a crucial role in a number of landmark discoveries in topological materials, including topological insulators and topological Dirac and Weyl semimetals. Here we review the latest developments in different types of ARPES techniques as well as some scientific applications in the study of topological materials. Lastly, we also discuss the future prospects of ARPES and highlight several promising applications on new topological materials.**

# Introduction:

Understanding collective behavior of electrons in solids has grown increasingly desirable as it is closely tied to many intriguing phenomena in condensed matter physics, such as superconductivity[1,2], quantum Hall effects[3–7], and etc. The nature of electrons in solids is described primarily by three quantum parameters: energy ($E$), momentum ($k$), and spin ($S$). Angle-resolved photoemission spectroscopy (ARPES)[8–10], due to its unique capability to directly probe the energy and momentum of electrons simultaneously, can therefore play a leading role in achieving a comprehensive understanding of the unique electronic properties of various types of solid-state materials.

Known generally as the photoelectric effect, the photoemission process was first discovered by Hertz in 1887[11], and its microscopic mechanism was explained by Einstein in 1905[12] when he introduced quanta of light - photon. A sketch of a typical ARPES measurement is shown in Fig. 1a, where a hemispherical photoelectron analyzer is used as an example. When light with sufficiently high frequency is incident on a sample, electrons can absorb photons and escape to the vacuum. These escaped electrons, known as photoelectrons, are then collected and analyzed with respect to their kinetic energy and outgoing angle by a spectrometer. Eventually, the energy and momentum of the electrons inside the sample are directly connected to the photoelectrons by conservation of energy and momentum parallel to the sample surface. Under the experimental configuration depicted in Fig. 1a, the following relationships hold:

$$E_{kin} = h\nu - \Phi - E_B \tag{1}$$

$$\hbar \vec{k}_\parallel^f = \hbar \vec{k}_\parallel^i = \sqrt{2mE_{kin}}\{\sin\theta\cos\varphi\hat{k}_x + \sin\theta\sin\varphi\hat{k}_y\} \tag{2}$$

where $E_{kin}$ is the kinetic energy of the photoelectrons, $E_B$ is the binding energy of the electron inside the sample, $\hbar k_\parallel^f$ and $\hbar k_\parallel^i$ are the parallel components (with respect to sample surface) of momenta of the photoelectron and the initial electron, respectively, θ and φ are the emission angles of the photoelectron, and Φ is the work function which is the energy required for an electron to escape from the Fermi level to the vacuum level, namely $\Phi = E_{vac} - E_F$ (Fig. 1b). Note that the above conservation laws are valid provided that the relaxation time of the electron-hole pairs is much longer than the escape time of the photoelectrons (~ tens of several attosecond)[13], and the momentum of the photon (~ 0.01 Å$^{-1}$ at 30 eV) is much smaller compared to the momentum of photoelectrons[8].

The momentum perpendicular to the surface, $k_\perp$, is not conserved since the surface of the material necessarily breaks the translational symmetry in this direction ($k_\perp^f \neq k_\perp^i$). However, $k_\perp^i$ can still be extracted by using a nearly free electron approximation for the final states, which is a reasonable approximation for a sufficiently high photon energy (usually tens of to hundreds of eV)[8,14–16]:

$$k_\perp^i = \sqrt{2m(E_{kin}\cos^2\theta + V_0)}/\hbar \qquad (3)$$

where $V_0$ is a constant called the inner potential. In practice, $V_0$ can be determined from the photon-energy-dependent measurements by fitting the experimental periodicity along the $k_\perp$ direction[14–17]. Once $V_0$ is determined, the value of $k_\perp^i$ could be extracted. Therefore, photon-energy-dependent ARPES measurement is an effective way to probe the electronic structure in the three-dimensional (3D) Brillouin zone, which is crucial for studying topological materials.

Strictly speaking, the photoemission process corresponds to a quantum many-body propagator describing the single-particle removal function from the full Fermi sea, which can be fully described by the so-called *one-step* model[8,9]. However, the one-step model is very complicated since it treats the photoemission as a single coherent process, which, in other words, describes the bulk, surface, and evanescent states within a single Hamiltonian. In many simplified cases, ARPES data are usually discussed within the *three-step* model[8,9], which separates the process into three sequential processes of photon absorption, electron travel to surface, and emission into vacuum.

In practice, ARPES intensity I($k$, $E$) can be written as:

$$I(k,E) = I_0(k,\nu,A)A(k,E)f(E,T) \qquad (4)$$

where A($k$, $E$) is the one-particle spectral function, from which we can extract the information about the quasiparticle self-energy that encodes the band structure and correlation effect. Here, f($E$,$T$) is the Fermi-Dirac distribution. The first part $I_0(k,\nu,A) \propto \sum_{f,i}|M_{f,i}^k|^2$, where $M_{f,i}^k = \langle \phi_f^k | A \cdot P | \phi_i^k \rangle$, is the photoemission matrix element, which describes the transition of the initial state $\phi_i^k$ to the final state $\phi_f^k$. ***P*** is the electronic momentum operator, and ***A*** is the electromagnetic vector potential, which depends on photon polarization and energy. The matrix element carries no direct information on the band dispersion. However, it can provide important orbital information of the electronic states if one implements some specific measurement geometries, which were widely used in studies of iron-based superconductors[18–20].

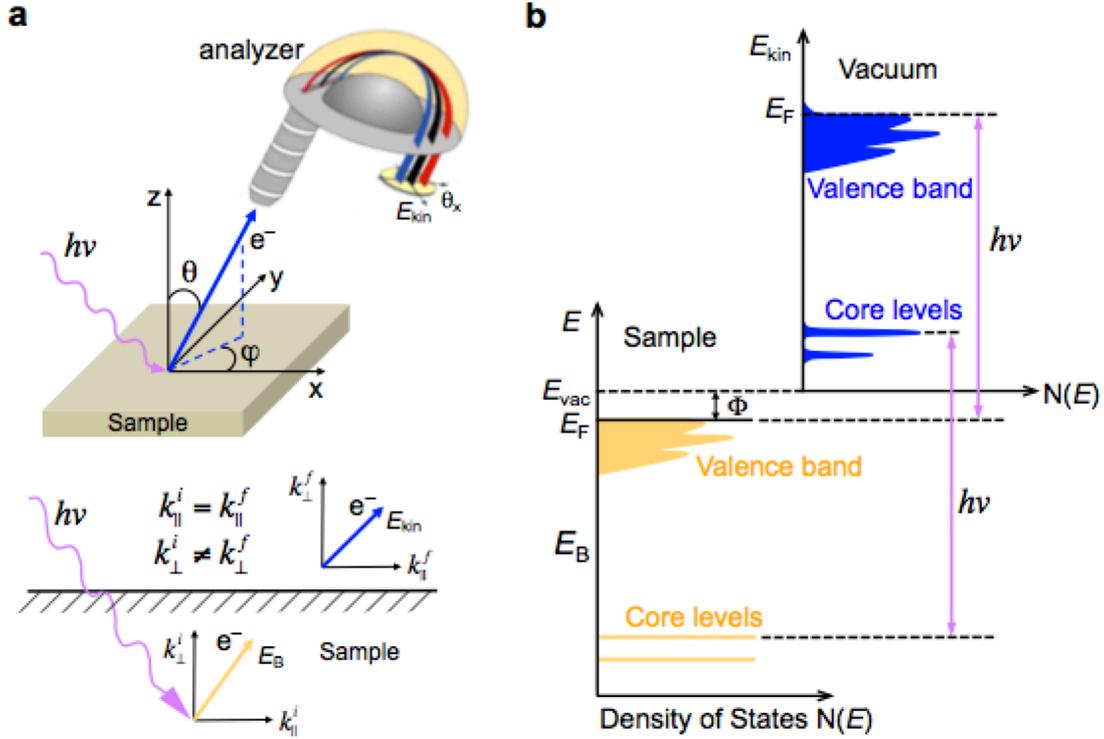

Figure 1 | Schematic diagram of ARPES measurements. **a** | Geometry and basic principles of ARPES measurements. Under incidence of light with photon energy $hv$, electrons are ejected from the surface of a sample at an angle ($\theta$, $\varphi$) and collected with an electron analyzer. In the photoemission process, the translational symmetry in the directions parallel to the sample surface is still preserved. Thus, the in-plane momentum is conserved, i.e., $k_{\parallel}^{i} = k_{\parallel}^{f}$. **b** | Energetics of the photoemission process, which implies the conservation of energy ($E_{kin} = hv - \Phi - E_B$). Panel **a** is adapted with permission from ref.[8], American Physical Society.

The basics of ARPES have been described in detail in a number of textbooks and review articles[8–10,21–25]. We briefly gave an introduction to some salient points of ARPES in the above paragraphs, which would be useful in understanding ARPES studies of topological materials. In the past decades, with the rapid development of different types of electron spectrometers[26–28], as well as modern synchrotron[29–35] and laser light sources[36–58], the ARPES technique has experienced a renaissance. The significantly improved energy and momentum resolutions[38,39] with laser light sources not only allow us to measure fine features of the electronic states, such as the superconducting gap in superconductors, with unprecedented precision[59], but also make it possible to obtain important information on many-body interactions in strongly correlated systems[60–62]. The use of continuously-tunable soft X-rays greatly enhances the bulk sensitivity of ARPES[17,29,33,63,64], which is crucial for studying the bulk electronic structure of 3D materials, especially for topological semimetals[65–69]. The integration of spin detectors into ARPES photoelectron spectrometers further extended the capability of ARPES[70–90], enabling quantification of the spin polarization of the band structures[91–95]. Moreover, the emergence of light sources with micron- or even nano-scale spot sizes have given rise to the possibility of performing spatially resolved ARPES measurements[31,96–100], which will play an irreplaceable role in probing the electronic structure of micro- and nano-scale materials as well as materials with phase separation or multiple domains[101–113]. Finally, the implementation of time-resolved ARPES with ultrafast lasers or X-ray sources[37,40–42,44–52,56,57] not only makes it possible to study ultrafast electronic dynamics in the time domain[114–157], but also enables us to probe the unoccupied states above the Fermi energy ($E_F$)[158–160]. These advances have enabled the application of ARPES in topological materials to aid our understanding of their unique band structures and physical properties.

3D topological materials are usually characterized by the nature of the surface states induced by the topology of the bulk band structure, which can be simply divided into two groups with respect to the bulk band gaps: insulators[91,92,161–171] with nonzero band gaps (e.g. topological insulators[91,92,161–166], topological crystalline insulators[167–171], etc.), and semimetals[65–69,172–197] with no band gaps (e.g. Dirac semimetals[172–176], Weyl semimetals[65,66,69,177–184], etc.). Taking advantage of different kinds of ARPES techniques such as vacuum ultraviolet (VUV) ARPES, soft x-ray ARPES, spin-resolved ARPES, etc., researchers have made a number of landmark discoveries in topological materials in recent years. In this topical review, we focus principally on the latest developments and future prospects of ARPES technique. To better illustrate the capability of this technique, we also highlight recent ARPES results from several topological materials, including the well-known topological insulator $Bi_2Se_3$ family[92,163,164] and the Weyl semimetal TaAs family[65,66,69,179–181].

**VUV ARPES**

An important parameter of ARPES is the incident photon energy. In the past few years, the energy range of photons has greatly expanded thanks to the development of laser and synchrotron light sources. At present, the incident light can vary from VUV light to soft and even hard X-rays. Among these, the most commonly used incident photon is VUV light, which we will review below.

Figure 2a shows the universal curve of a photoelectron's inelastic mean free path as a function of its kinetic energy[198]. For incident photon energy above 20 eV in the VUV region, the short mean free path of photoelectrons makes ARPES an extremely surface sensitive technique. This means that a considerable fraction of the total ARPES signal will be representative of the topmost surface layer, which can be regarded as an advantage when dealing with surface states as in topological insulators and Weyl semimetals. Consequently, ARPES experiments have to be performed on atomically clean and well-ordered flat surfaces. To get a clean surface and to avoid surface contamination, single crystals are typically cleaved *in situ* and measured in ultra-high vacuum chambers. Even then, the freshly cleaved surface has a finite measurement lifetime. Typically, to maximize the sample measurement time, one should use a vacuum better than $5 \times 10^{-11}$ torr, which can be routinely achieved by using modern ultra-high vacuum systems. In many materials, flat mirror-like surfaces can be obtained by a straightforward cleaving. However, there still exists a large class of materials that are not cleavable, especially for 3D materials, which limits the applicability of ARPES. Alternatively, fresh thin films can be used, and it has now become routine to grow films in vacuum and transfer them immediately to an ARPES chamber after growth. Indeed, many combined ARPES systems have been developed in the recent years, with thin film growth capabilities such as molecular beam epitaxy (MBE) and pulsed laser deposition (PLD); numerous achievements have been made based on these combined systems[62,95,199–201]. Furthermore, we also note another powerful way to prepare atomically clean and flat surfaces by polishing, repeatedly sputtering and annealing the samples, known as the "polish-sputter-anneal" method[202]. This method is suitable for some materials that have strong chemical bonding and can be crystalized by annealing in vacuum. Very recently, this method has been successfully applied to the study of 3D topological semimetal CoSi family[203,204] and enables the observation of multiple types of unconventional chiral fermions and helical Fermi arc surface states on the (001) surface.

For VUV ARPES, there are three types of light sources that have been successfully used: noble gas discharge lamps[205–207], synchrotron radiation[29–35], and laser[36–58] light sources. At the current stage, these three VUV light sources are complementary for ARPES experiments, and their main features are summarized in Table 1. In the following we discuss them in detail.

*Table 1: Comparison of the three types of VUV light.*

| VUV Light source | Laser | Synchrotron | Discharge lamp |
|---|---|---|---|
| Photon energy | Discrete or tunable within a limited range 6 eV, 7 eV, 11 eV, etc. | Continuously tunable several eV - thousands of eV | Discrete 21.2 eV, 40.8 eV, etc. |
| Best energy resolution | < 1 meV[38,39] | 1-30 meV | 1.2 meV |
| Photon flux (photon/s) | ~ $10^9$-$10^{15}$ | ~ $10^{12}$-$10^{13}$ | ~ $10^{12}$ |
| Spot size | ~ 100 μm | Tens of nm[98] – hundreds of μm | ~ 1 mm |
| Polarization | Tunable | Tunable | Non-tunable |
| Detection range | Limited | Several Brillouin zones | First to second Brillouin zones |

*Lamp-based VUV ARPES*

As the earliest photoemission light sources, noble gas (helium, neon, argon, krypton, and xenon) discharge lamps have been used widely in laboratory-based ARPES systems. These lamps utilize gas resonance lines as monochromatic light sources. A typical and commonly used example is the modern helium lamp, which employs the microwave plasma technique[206]. Such a helium lamp can mainly provide two discrete photon energies, 21.2 eV (He Iα) and 40.8 eV (He IIα), which can be selectively used by adjusting the angle of a monochromator. Additionally, the monochromator can refocus the beam to a spot size of ~ 1 mm with a photon flux of ~ $10^{12}$ photon/s.

In comparison with the other two VUV sources (laser and synchrotron radiation), discharge lamps are compact, stable, and particularly affordable to most laboratories. More importantly, the photon energy provided by discharge lamps is suitable for typical ARPES experiments due to reasonable cross sections[8,9,208] and a negligible space charge effect[209–211]. However, discharge lamps have some obvious limitations: relatively low photon flux especially for high-resolution measurements, non-tunable photon energy with fixed or no polarization, relatively low momentum resolution, a requirement of very flat surfaces due to the large beam spot size, and spillage of noble gas into the ARPES chamber during measurements. Nevertheless, noble gas discharge lamps are still the most popular and favorable lab-based light sources, which are very useful in the study of two-dimensional materials and thin films[199–201].

### Synchrotron-based VUV ARPES

When electrons move at a relativistic velocity (close to the speed of light) and are bent by a static magnetic field, there will be electromagnetic radiation from the electrons due to radial component of acceleration. This radiation, known more widely as synchrotron radiation light, can be applied to ARPES once properly monochromatized. Indeed, benefitting from the continuous development of synchrotron technology, especially the advent of third generation synchrotron light sources, many synchrotron-based ARPES end stations have been successfully built and have gradually become the most powerful ARPES systems in the past decades[29–35].

Synchrotron-based ARPES features many distinctive advantages. First, taking advantage of the variable polarization undulator and high-resolution monochromator, one can easily tune the photon energy and polarization (linear or circular) of the beam. The continuously tunable photon energy not only allows one to map out the electronic structure in the entire 3D momentum space, but also makes it possible to distinguish surface and bulk states by performing photon-energy-dependent ARPES measurements[65,66,164,181]. Moreover, by conducting polarization-dependent measurements, one can also identify the orbital character of the bands based on matrix element effects[18–20]. Additionally, within the focusing system and tunable beam slit, a high-flux beam (~ $10^{13}$ photon/s) with a sub-100 μm spot size can routinely be achieved, thus facilitating ARPES measurements on relatively small samples.

Despite these amazing capabilities, synchrotron-based ARPES has its limitations. One main concern is the high cost and significant construction and maintenance efforts of the synchrotron light sources. Moreover, it remains a challenge to achieve super-high energy resolution (< 1 meV) and high-flux synchrotron light since a decrease in the beam bandwidth has to be at the sacrifice of photon flux.

Most synchrotron-based ARPES setups are equipped with continuously tunable VUV light, with which many important breakthroughs have been made in the study of topological materials. Here we will give an example: the observation of surface Fermi arcs in the Weyl semimetal TaAs[69,181]. Weyl semimetals[65,66,69,177–184] are a class of materials that can be regarded as 3D analogs of graphene wherein the bulk non-degenerate electronic bands disperse linearly along all momentum directions through a node near $E_F$, called a Weyl node, which can be viewed as a singular point of Berry curvature or "magnetic monopole" in the momentum space[212]. Weyl nodes can only appear in pairs of opposite chirality in a real material when the spin-doublet degeneracy of the bands is removed by breaking either time-reversal or inversion symmetry[177–180]. Interestingly, the low-energy excitations near the Weyl nodes behave as Weyl fermions, which were originally proposed by Hermann Weyl in 1929 as massless solutions of the Dirac equation[213]. However, their existence in particle physics has not yet been observed after more than eight decades, which makes the physical realization of Weyl fermions in Weyl semimetals more significant.

One hallmark of a Weyl semimetal is the existence of pairs of entangled Fermi arcs at two opposite surfaces connecting the projections of two Weyl nodes with opposite chiralities[177], as sketched in Fig. 2b. Synchrotron-based VUV ARPES, with its unique tunable light, has played a central role in identifying surface Fermi arcs in the TaAs family (TaAs, TaP, NbAs, NbP). As an example, by performing photon-energy-dependent measurements along the Γ-Y direction in TaAs (Fig. 2d), the surface states at $k_y$ ~ 0.4 Å$^{-1}$ are clearly identified since they do not show any noticeable change by varying the incident photon energy[69]. To identify the surface Fermi arcs, one should then map out the Fermi surfaces. Indeed, the surface Fermi arc

a5 is clearly resolved in the Fermi surface map at hv = 20 eV (Fig. 2e), which matches well with calculations[181] (Fig. 2c).

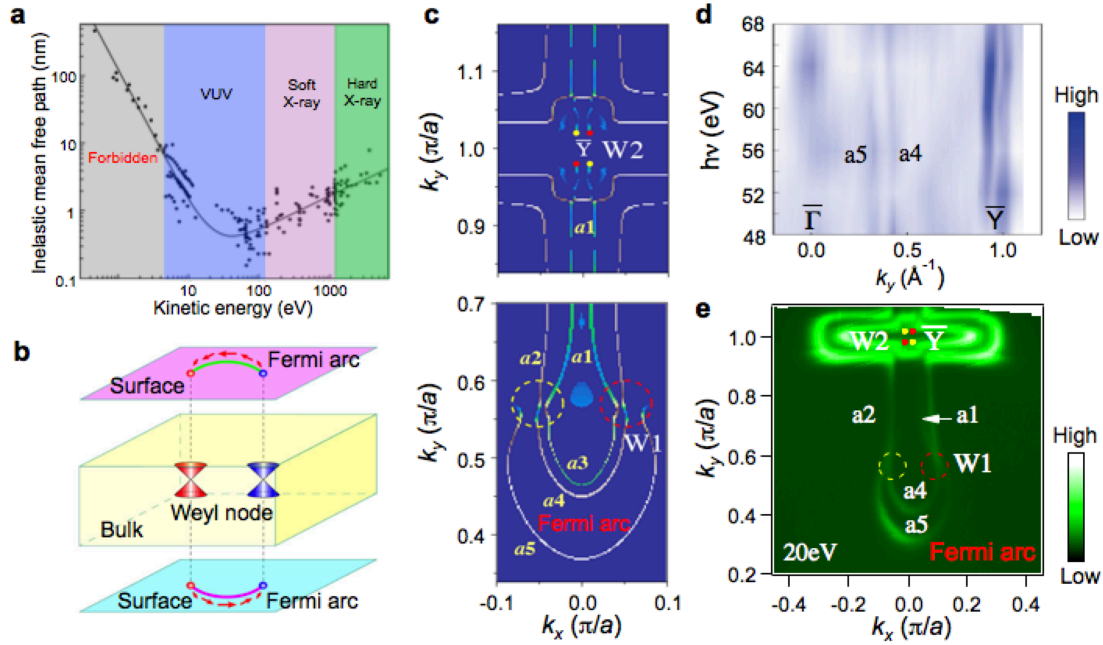

Figure 2 | Surface Fermi arcs on the TaAs (001) surface measured by synchrotron-based VUV ARPES. **a** | Universal curve of electron mean free path as a function of kinetic energy. **b** | Schematic of a Weyl semimetal with spin-polarized Fermi arcs on its surfaces connecting the projections of two Weyl nodes with opposite chiralities (indicated by red and blue colors). The red arrows on the surfaces indicate the spin texture of the Fermi arcs. **c** | Fine calculations of surface states at $E_F$ around Weyl points W2 (circular dots) and W1 along $\overline{\Gamma}-\overline{Y}$ direction, respectively. Weyl points W1 are indicated by dashed circles since the chemical potential is slightly away from the nodes. Yellow and red colors represent opposite chiralities. **d** | ARPES intensity plot at $E_F$ along $\overline{\Gamma}-\overline{Y}$ collected at different photon energies (48-68 eV). **e** | ARPES intensity plot at $E_F$ recorded on the (001) surface at $hv$ = 20 eV. Panel **a** is adapted with permission from ref.[198], John Wiley and Sons. Panel **b** is adapted with permission from ref.[93], American Physical Society. Panel **c** is adapted with permission from ref.[181], CC-BY-3.0. Panel **d** is adapted from ref.[69], Springer Nature Limited.

### *Laser-based VUV ARPES*

Without considering the photon energy, lasers seem to be a natural choice for laboratory-based ARPES due to their high photon flux and narrow bandwidth. However, the photon energies generated by lasers are usually too low to overcome the 4–5 eV work function of the material under an ARPES measurement. Great efforts have been made in recent years to develop high-energy (> 5 eV) lasers that are suitable for ARPES. Fortunately, taking advantage of nonlinear optical processes, long sought VUV lasers have been achieved[36–58].

In general, the VUV laser light required for the photoemission process can be generated in two different ways. One way is using harmonic generations in nonlinear optical crystals. The most commonly used nonlinear optical crystals are $BaB_2O_4$ (BBO) and $KBe_2BO_3F_2$ (KBBF). By using BBO crystals, a Ti:sapphire laser (~ 800 nm) can output ~ 6 eV laser light[36], whereas with KBBF crystals, a $Nd:YVO_3$ laser (~ 1064 nm) can produce ~ 7 eV laser light with an ultra-narrow bandwidth and a high flux[38,39]. The other way to generate VUV laser light is based on high harmonic generation (HHG) or multi-photon excitations in the noble gases[41,42,46,53,214], which can generate higher photon energy laser light compared to nonlinear optical crystals. For example, a fiber laser with a Xe gas cell can output ~ 11 eV laser light[53], and a Ti:sapphire laser with gas cell can even generate 15–40 eV tunable laser light with attosecond[57] or femtosecond pulse duration[46]. However, the HHG sources usually have relatively low photon flux ($\leq 10^{13}$ photon/s) due to the low generation efficiency of the HHG process and unavoidable loss in the optics after generation. Although HHG laser light is not suitable for high-resolution ARPES measurements owing to relatively low energy resolution (~ tens of meV) that is limited by the Heisenberg uncertainty principle, such femtosecond or attosecond-scale pulsed laser light makes it possible to perform time-resolved ARPES measurements at relatively high photon energies.

Laser-based ARPES has many unique advantages. One major benefit of laser ARPES with relatively low-energy (~ 6–7 eV) photons is a significant gain in in-plane momentum ($k_\parallel^f$) resolution. This is because $\Delta k_\parallel^f$ is proportional to $\sqrt{2mE_{kin}/\hbar^2} \cdot \cos\theta \cdot \Delta\theta$ from Eq. (2); therefore, at the same $\theta$ and $\Delta\theta$, low kinetic energy photoelectrons will result in a better in-plane momentum resolution. Moreover, the high photon flux (~ $10^{15}$ photon/s) and extremely narrow bandwidth (< 1 meV) of the 6-7 eV laser light also makes it possible to perform super-high energy resolution ARPES measurements with sufficient data acquisition efficiency. Thus, the relatively low-energy laser light sources have the advantage of ultra-high resolutions in both energy and momentum. Indeed, laser-based ARPES systems with a photon energy of ~ 7 eV and energy resolution better than 1 meV have been developed by the harmonic process using a KBBF nonlinear crystal[38,39]. In addition, unlike noble gas discharge lamps, it is easy to control the polarization of the laser light and to tune the beam spot size to a micro-scale. More importantly, pulsed laser light adds a new degree of freedom to ARPES - time resolution- which we will discuss in more detail later.

As with the other VUV light sources, laser light also has its weaknesses. The major disadvantages are limited photon energy tunability, poor cross-section for some materials, and the inability to access electrons far from the Brillouin zone center. In some sense, the application of HHG laser light with tunable photon energies of tens of eV has successfully broken the above barriers. However, it is still difficult to get high-stability and high-flux photons with a small bandwidth.

To elucidate the high-resolution power of laser ARPES, we highlight one application: the observation of a topological surface Dirac cone on the (001) surface of $FeTe_{0.55}Se_{0.45}$[215]. As the simplest iron-based superconductor (Fig. 3a), Fe(Te,Se) has attracted tremendous attention since its discovery[216]. And recently, calculations[217–219] also predicted the existence of topological order in $FeTe_{0.55}Se_{0.45}$. As shown in Fig. 3b, the substitution of Se with Te shifts the $p_z$ band down and leads to a band inversion between the $p_z$(-) and $d_{xz}/d_{yz}$(+) bands along the Γ-Z direction. Eventually, owing to the opening of the spin-orbit coupling (SOC) gap between these two inverted bands, this material naturally hosts strong topological surface states inside the gap at the $\bar{\Gamma}$ point of the (001) surface (Fig. 3c). Unfortunately, limited by energy and momentum resolutions, former ARPES measurements could not clearly resolve these surface states[220,221]. However, very recently, Zhang *et al*. revealed the Dirac surface states by taking advantage of high-resolution laser ARPES[215]. Figure 3 summarizes their key ARPES evidence, and the Dirac-cone surface state near $E_F$ is clearly resolved from the high-resolution curvature intensity plot (Fig. 3d). More importantly, when $FeTe_{0.55}Se_{0.45}$ enters the superconducting state, superconductivity will also be induced on the topological surface states, as illustrated in Fig. 3e. Figures 3f-h show the temperature- and momentum-dependent measurement of the induced superconducting gap (recorded at 2.4 K). Remarkably, the energy distribution curves at slightly higher temperatures show evidence of the hole branch of Bogoliubov quasiparticles (the shoulders above $E_F$ in Fig. 3f), demonstrating its superconducting nature. In addition, the superconducting gap on the topological surface state is also isotropic (Fig. 3g, h), which is the same as in the Fu-Kane model[222] and thus can support Majorana zero modes in its vortex cores[223].

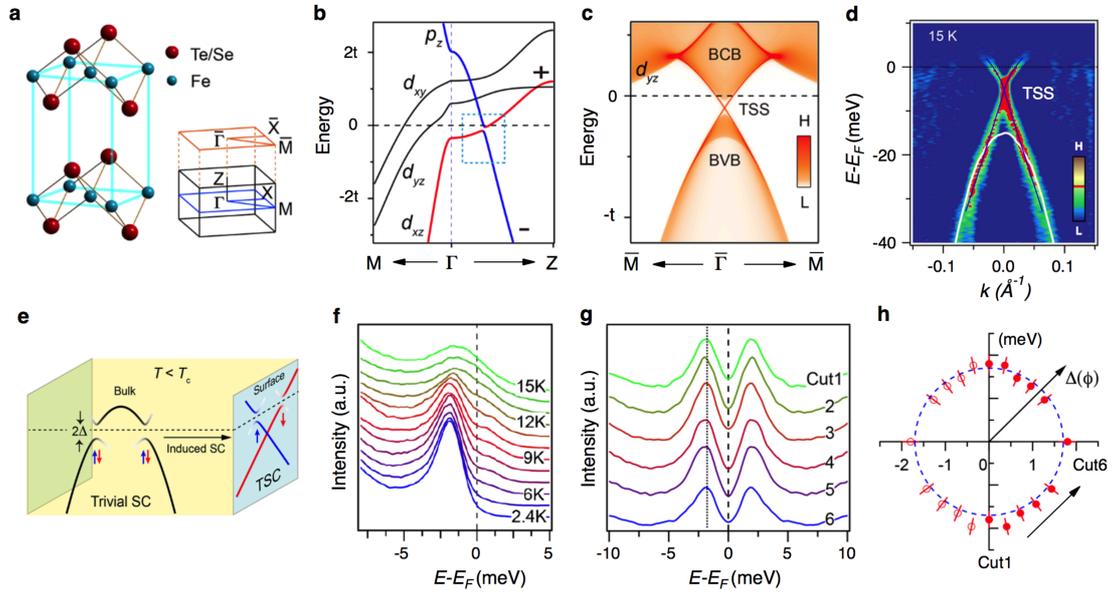

Figure 3 | Topological surface Dirac cone on the (001) surface of FeTe$_{0.55}$Se$_{0.45}$ measured with laser-based VUV ARPES. **a** | Crystal structure, and bulk and projected (001) surface Brillouin zones of FeTe/Se. **b** | Calculated band structures of FeTe$_{0.55}$Se$_{0.45}$ along the Γ-M and Γ-Z directions. The dashed box shows the SOC gap of the inverted bands. **c, d** | Calculated band structure (**c**) and curvature intensity plot of ARPES data (**d**) along the $\overline{\Gamma} - \overline{M}$ direction. The topological surface states (TSS) connecting the bulk valence band (BVB) and bulk conduction band (BCB) are clearly resolved in both **c** and **d**. The ARPES data in **d** is recorded with a *p*-polarized 7-eV laser. **e** | Schematic of the bulk and surface superconducting (SC) states in FeTe$_{0.55}$Se$_{0.45}$. Below $T_c$, the bulk states open *s*-wave SC gaps, which are topologically trivial because of their spin degeneracy (black curves). Induced by the bulk-to-surface proximity, the TSS also opens an *s*-wave gap at $T < T_c$, and are topologically superconducting (TSC) as a consequence of the spin polarization (blue and red curves). **f** | Raw energy distribution curves measured at different temperatures for a *k* point on the surface Fermi surface. The shoulders above $E_F$ signify the SC Bogoliubov quasiparticles. **g** | Symmetrized energy distribution curves of the Dirac surface states at different Fermi wave vectors (indicated in **g**) recorded at $T = 2.4$ K. **h** | Polar representation of the measured SC gap size in **f**. The hollow markers are a mirror of the solid markers. Figure is adapted with permission from ref.[215], AAAS.

## Bulk sensitive soft X-ray ARPES

Synchrotron-based VUV ARPES, usually with hv ≥ 20 eV, is an ideal tool for studying the surface electronic structure of many materials, especially topological materials. However, it is difficult to probe the bulk states of 3D materials with the above 20eV VUV light due to the extremely short escape depths of photoelectrons (Fig. 2a) and the ill-defined surface-perpendicular wavevector $k_\perp$.

To examine the 3D bulk states, especially the band dispersion along the $k_\perp$ direction, the only reliable way is to go to higher photon energies in the soft X-ray regime[17,29,33,63,64]. We explain the reasons below: (i) the increase of the photoelectron kinetic energy by soft X-rays results in an increase of the photoelectron escape depths by a factor of two to four compared with 20 eV VUV light (~ 5 Å), which greatly enhances the bulk or buried layer/interface sensitivity, (ii) from the Heisenberg uncertainty principle, the intrinsic $k_\perp$ broadening of photoelectrons can be defined by $\Delta k_\perp \sim d^{-1}$, where $d$ corresponds to the photoelectron escape depth. Thus, the increase of $d$ by soft X-rays would naturally lead to the improvement of the intrinsic $k_\perp$ resolution, which enables accurate investigations of the bulk states of 3D materials in the whole momentum space, and (iii) in the photoemission process, high-energy soft X-ray incident light makes the final states truly free-electron-like, which allows precise determination of the $k_\perp$ value from Eq. (3). Notably, the low-energy laser light also has a longer penetration depth (Fig. 2a). However, the non-tunable photon energy and stronger final state effects[69] due to low-speed photoelectrons makes it nearly impossible to probe the intrinsic 3D band structure of bulk states.

Besides the bulk or buried layer/interface sensitivity, soft X-ray ARPES also features other advantages, such as simplified matrix elements and relatively insensitive to sample surface quality compared with VUV ARPES. However, the soft X-ray ARPES also suffers from several problems. The main difficulty is a decrease of the valence band cross section by two to three orders of magnitude compared to VUV energy range[208], due to the fact that the overlap of the rapidly oscillating high-energy final states with the smooth valence states is reduced to the small ion core region[224]. Such a dramatic signal loss has to be compensated by high flux of incident photons. Indeed, thanks to the advancement of synchrotron radiation sources and beamline instrumentation, the soft X-ray ARPES endstations at Swiss Light Source (SLS) has successfully broke this barrier with high photon flux (> $10^{13}$ photon/s)[33]. Another difficulty with soft X-ray ARPES, in sharp contrast to low-energy laser ARPES, is that the use of high-energy soft X-rays would sacrifice the in-plane momentum resolution. There is therefore a need for higher angular resolution spectrometers to improve the $k_\parallel$ resolution. Furthermore, compared to VUV ARPES, soft X-ray ARPES also suffers from a loss of energy resolution. More specifically, the energy resolution of VUV ARPES can be better than 1 meV, while for soft X-ray ARPES, the energy resolution varies from tens of meV to a hundred meV depending on the photon energy. Lastly, we caution that the photon momentum of soft X-rays may not be negligible during the photoemission process, which complicates the momentum conservation law.

To illustrate how one can study the bulk electronic structures by soft X-ray ARPES, and how critical the improvement in $k_\perp$ resolution has been in this regard, we use 3D topological semimetals TaAs[65,66] and MoP[67] as examples. Topological semimetals[65–69,172–197] with symmetry-protected band-crossing points have grown as one of the most intensively studied fields in current condensed matter physics. The most famous examples are Dirac[172–176] and Weyl semimetals[65,66,69,177–184], in which two doubly or singly degenerate bands cross each other, forming four-fold Dirac points or two-fold Weyl points (Fig. 4a). Recently, theorists predicted new types of crystal symmetry-protected band crossings in condensed matter systems and the corresponding low-energy excitations have no high-energy counterparts[185–193]. For example, topological semimetals with 3-fold band crossings have been predicted to exist in several materials with tungsten carbide structures[188–190]. The quasiparticle excitations near the band-crossing point are three-component fermions, which can be viewed as the "intermediate species" between the four-component Dirac and two-component Weyl fermions.

With the ability to probe more deeply into a sample, soft X-ray ARPES has played a key role in detecting the bulk Weyl points in TaAs[65,66] and triply degenerate points in MoP[67] and WC[68]. The benefits of this are clearly illustrated in Figs. 4c-d. With regard to the Weyl semimetal TaAs, first-principles calculations have predicted that there exist 12 pairs of Weyl points in the bulk Brillouin zone[179] (Fig. 4b). Unfortunately, due to the short escape depth of the photoelectrons excited by VUV light, the bulk Weyl points were not resolved in the VUV ARPES experiments[181] (Fig. 2e). However, with soft X-ray ARPES the bulk Weyl bands are clearly resolved[65,66]. As shown in Fig. 4c, the measured electronic states in the Γ-Σ-Z-S plane, which is perpendicular to the cleaved sample surface, clearly exhibit a periodic modulation upon varying the incident soft X-ray photon energy ($k_\perp$), confirming the bulk nature of the detected spectra. The W1 bulk Weyl points are confirmed from the measured "M"-shaped band dispersion in Fig. 4d.

The benefit of the soft X-ray ARPES technique is further illustrated in Figs. 4e-i, which show ARPES results from a MoP sample. In MoP, due to the 3D crystal structure, the cleaved surface is disordered (Fig. 4e). Consequently, at 60 eV, no obvious Fermi surface is seen (Fig. 4f), due to a combination of the surface sensitivity of VUV ARPES and the angle-smearing effect of elastic scattering in the amorphous layer. However, upon increasing the energy to 453 eV, the bulk states are clearly seen (Fig. 4g), which enables the detection of the predicted triply degenerate point. Indeed, by using high-precision measurements of the band dispersions, we clearly resolve the triply degenerate point with soft X-rays (Fig. 4h), which is well reproduced by the calculations in Fig. 4i.

In summary, the state-of-the-art soft X-ray ARPES technique features a heightened sensitivity to bulk or buried layers/interfaces as demonstrated by the two aforementioned examples. This makes it a powerful technique to reveal the bulk band crossings of 3D topological semimetals.

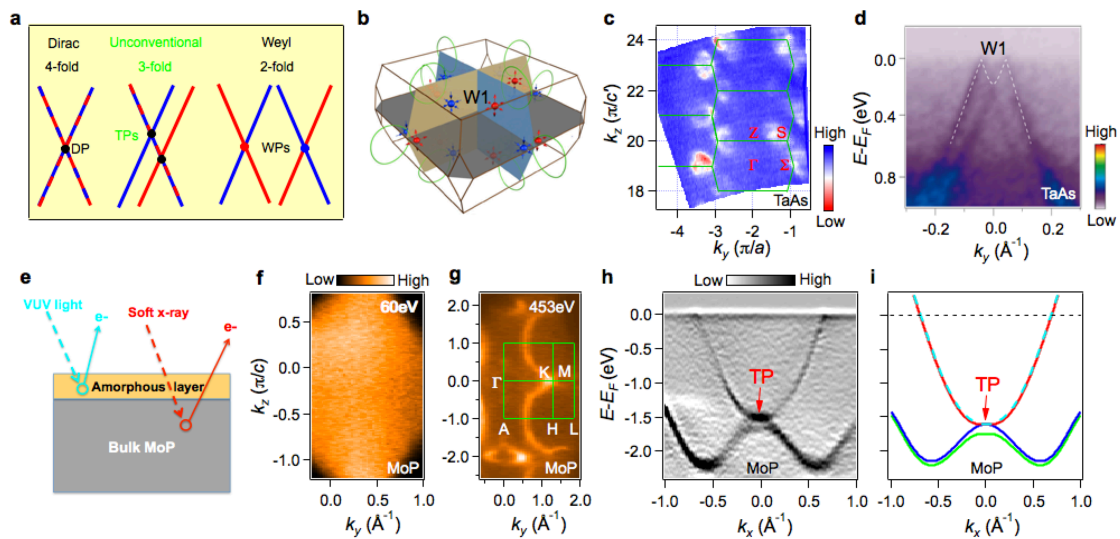

Figure 4 | Bulk-sensitive soft X-ray ARPES studies of the bulk Weyl points and triple points in TaAs and MoP, respectively. **a** | Schematics of Dirac fermions, unconventional three-component fermions,

and Weyl fermions, which have band crossing points with four- (Dirac point, DP), three- (triply degenerate point, TP) and two-fold (Weyl) degeneracies, respectively. **b** | 3D view of the nodal rings (in the absence of SOC) and Weyl points (with SOC) in the bulk Brillouin zone of TaAs. **c** | ARPES intensity plot at $E_F$ in the $k_y$-$k_z$ plane at $k_x$ =0. The green lines represent the Brillouin zone structure in the $k_y$-$k_z$ plane. c' is one half of the *c*-axis lattice constant of TaAs. **d** | ARPES intensity plot showing experimental `M'-shaped band dispersions of one pair of Weyl nodes (W1) in TaAs. **e** | Schematic photoemission process with bulk-sensitive soft X-rays and surface-sensitive VUV light on the (100) cleavage surface covered by an amorphous layer for MoP single crystal. **f, g** | ARPES intensity maps in the $k_x$ = 0 plane at $E_F$ recorded on the same (100) cleaved surface of MoP with soft X-ray (453 eV) and VUV (60 eV) light, respectively. **h, i** | Curvature intensity plot of ARPES data and calculated band structure along the $k_x$ direction at $k_z$ = 0.75π (correspond to the TP in MoP). Panels **a**, **g**, **h** and **i** are adapted from ref.[67], Springer Nature Limited. Panel **b** is adapted with permission from ref.[179], CC-BY-3.0. Panel **c** is adapted from ref.[65], Springer Nature Limited. Panels **d** is adapted from ref.[66], Springer Nature Limited.

## Spin-resolved ARPES

Historically, development of the ARPES technique has been heavily driven by increasing research demand. For example, the discovery of the high-$T_c$ superconductors promoted the development of high-resolution laser ARPES[38,39]. Similarly, the recent discovery of topological insulators as well as non-centrosymmetric materials has stimulated the development of high-performance spin-resolved ARPES[70–90].

The integration of spin detectors to ARPES spectrometers enables spin-resolved ARPES. Many efforts have been made to develop compact spin polarimeters, including spin-polarized low-energy electron diffraction[225], diffuse scattering[226], Mott scattering[73], and very low energy electron diffraction (VLEED) [227]. Most of these spin detectors are based on the asymmetry (A) of preferential spin scattering, which can be written as: $A = (I_+ - I_-)/(I_+ + I_-)$, where $I_+$ and $I_-$ are the intensity of electrons detected by the parallel and anti-parallel channels, respectively. The final spin polarization rate (*P*) is proportional to the asymmetry: *P* = A/S, where S is the spin sensitivity (a coefficient called the Sherman function with Mott detectors), which can be determined by measuring a fully polarized electron beam with $P = 1$. The most widely used spin polarimeters are Mott detectors[73] and VLEED detectors[227], which utilize heavy elements, such as Au and Th, and ferromagnetic thin films, such as Fe (001) and Co(001) films, as scattering targets, respectively. The Mott detector utilizes the spin-orbit interaction-induced scattering asymmetry when the high-energy (~ 20-100 keV) electrons scatter off a heavy-element target. By contrast, the VLEED detector takes advantage of the exchange scattering asymmetry of very low-energy (below 10 eV) electrons with ferromagnetic thin films. Mott detectors have the advantage of high stability and high accuracy, but their scattering rate is very low (~ $10^{-4}$ efficiency compared to spin-integrated ARPES) due to low cross sections of high kinetic energy electrons. On the other hand, VLEED detectors have much higher efficiency (~ two orders of magnitude higher than Mott detectors) thanks to the higher scattering probability of very low-energy electrons and improved spin sensitivity, but the ferromagnetic thin-film target has to be regenerated frequently, typically once every few weeks, due to degradation of the thin film.

Spin-resolved ARPES is one of the most powerful techniques that can directly measure the spin texture of electronic states. Similar to soft X-ray ARPES, spin-resolved ARPES also suffers from several shortcomings, including low efficiency due to the scattering process and single-detection channel of spin detector, and relatively low energy and angular resolutions. However, recent advances in spin detectors[71,76–80,84] and light sources have greatly increased the performance of spin-resolved ARPES. For example, combining a DA30-L spin spectrometer and a high-photon-flux VUV laser, a high efficiency spin-resolved ARPES system with an energy resolution of 1.7 meV was achieved at Tokyo University[85]. Fig. 5a shows the schematic of the DA30-L spin spectrometer. The spin spectrometer is a combination of ScientaOmicron DA30-L hemispherical electron analyzer and twin VLEED spin detectors (VLEED-B and VLEED-W). The DA30-L analyzer is equipped with electron deflectors, whereby an applied electric field is used to select photoelectrons of desired emission angles. Therefore, the DA30-L analyzer allows detailed k-space mapping of two dimensional (2D) in-plane electronic structure, $E(k_x,k_y)$, at a fixed sample geometry with acceptance angles $\theta_x$ (direction along the entrance slit) and $\theta_y$ (direction perpendicular the entrance slit) of 30° and 24°, respectively. The twin VLEED spin detectors utilize oxidized iron films as scattering targets, and the film target of VLEED-B (VLEED-W) is magnetized by magnetic coils, in the $x$ and $z$ ($y$ and $z$) directions, which correspond to the spin polarization directions of $P_x$ and $P_z$ ($P_y$ and $P_z$) on the sample axis. Thus, the DA30-L spin spectrometer enables us to analyze the 3D spin vector of the photoelectrons emitted in the acceptance cone of ($\theta_x \times \theta_y$) = (30°×24°) without sample rotation. Lastly, we caution that the spin-polarization signal can be complicated by matrix element effects[228–230]. Thus, photon-energy-dependent and polarization-dependent spin-resolved ARPES measurements are needed to check whether the spin signal is intrinsic[93,94].

As mentioned previously, the discovery of 3D topological insulators greatly promoted the development of spin-resolved ARPES. A topological insulator is a novel state of quantum matter, which features an energy gap in the bulk but gapless Dirac-cone surface states that reside inside the bulk insulating gap[91,92,161,164–166]. The most distinct feature of a Dirac-cone surface state is spin-momentum locking pattern[165], which manifests itself as a spin texture that winds in a circle around a constant-energy contour of the Dirac-cone surface state (Fig. 5b). To examine the helical nature of the surface states, we use $Bi_2Te_3$ as an example since it hosts a clean Dirac cone near $E_F$ (Fig. 5c). Indeed, Hsieh *et al.* observed the spin-momentum-locking feature of the surface electrons in $Bi_2Te_3$ with spin-resolved ARPES[91]. Fig. 5d,e shows the measured spin-polarization spectra along $\overline{\Gamma} - \overline{M}$ in the $x$, $y$, and $z$ directions, respectively. While no clear spin polarization is observed in the $x$ and $z$ directions within the experimental resolution, clear polarization signals of equal magnitude and opposite signs are observed in the $y$ direction. This implies that the surface electrons of opposite momenta also have opposite spin textures, confirming the spin-momentum-locking scenario. Note that besides the spin-resolved ARPES, the spin textures of electronic states can also be extracted based on the spin-dependent differential absorption of left- versus right- circularly polarized light, known as circular dichroism[231–233].

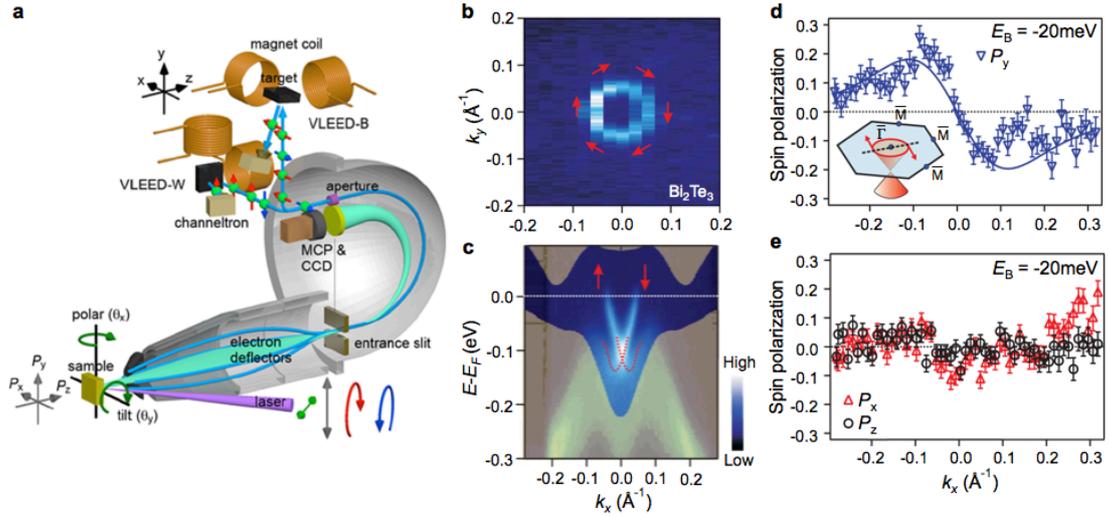

Figure 5 | Spin texture of surface Dirac cone of $Bi_2Te_3$ measured with spin-resolved ARPES. **a** | Schematic of the spin-resolved ARPES system from Prof. Shin's group at Tokyo University. Twin VLEED spin detectors arranged with orthogonal geometry are connected to the end of DA30-L hemispherical electron analyzer. **b** | ARPES intensity map at $E_F$ recorded on the (111) cleaved surface of $Bi_2Te_3$. Red arrows denote the direction of spin projection around the Fermi surface. **c** | ARPES intensity plot along the $k_x$ ($\overline{\Gamma}-\overline{M}$) direction. The dotted red lines are guide to the eye. The shaded regions are the calculated projections of the bulk bands onto the (111) surface. **d** | Measured $y$ component of spin polarization along the $\overline{\Gamma}-\overline{M}$ direction at a binding energy of 20 meV, which only cuts through the surface states. Blue solid lines are numerical fits, the inset shows a schematic of the cut direction as well as corresponding spin texture. **f** | Measured $x$ (red triangles) and $z$ (black circles) components of spin polarization along the $\overline{\Gamma}-\overline{M}$ direction. Panel **a** is adapted with permission from ref.[85], American Institute of Physics. Panels **b-e** are adapted from ref.[91], Springer Nature Limited.

## Ultrafast time-resolved ARPES

Conventional ARPES serves as an excellent and reliable tool to probe the band dispersion of occupied states. However, it remains challenging to distinguish and quantify many-body interactions (e.g., electron-electron, electron-phonon interactions) of correlated materials with conventional ARPES. Generally, the electron-electron, electron-spin, and electron-phonon interactions occur in the femtosecond, tens of femtosecond, to tens of picosecond time scales[234–236]. Therefore, the use of sub-picosecond or even sub-femtosecond laser pulses in pump-probe experiments enables one to disentangle the coupled interactions between the charge, spin, lattice and orbital degrees of freedom[157]. In particular, the combination of pump-probe optical spectroscopy with ARPES, namely time-resolved ARPES[37,40–42,44–52], provides direct insights to the energy and momentum dependence of these ultrafast dynamics.

In a time-resolved ARPES experiment (Fig. 6a), a femtosecond laser pulse ("pump") is used to perturb a material into a non-equilibrium state. Subsequently, a second time-delayed pulse ("probe"), typically ultraviolet, is used to excite photoelectrons out of the sample, which are analyzed by an electron spectrometer. Varying the time delay between the pump and probe pulses makes it possible to obtain insight into the time-dependent processes involved in the relaxation of these transient states. Besides providing energy- and momentum-resolved transient spectra in the time domain[114–157], time-resolved ARPES also allows for the investigation of unoccupied states above $E_F$ by occupying them by photo-excited electrons while keeping the band structure minimally disturbed[158–160].

Typically, femtosecond VUV laser pulse generated from nonlinear crystals or noble gases is used as the probe photon source. For the pump pulse, a variety of frequencies, ranging from visible light down to terahertz light[154], are applicable, among which the most commonly used photon energy is 1.55 eV, since 1.55 eV femtosecond laser pulses can easily be generated with a Ti:sapphire laser. The advances and limitations of VUV laser ARPES have been discussed above. Here we further note that the implementation of ultrashort femtosecond pulses in time-resolved ARPES measurements results in limited energy resolution due to the Heisenberg uncertainty principle. Another drawback is the low efficiency ($\leq 10^{-1}$ efficiency compared to conventional ARPES) due to relatively low intensity of laser pulses constrained by the space charge effects, as well as the finite lifetime of the transient states. Consequently, a high repetition rate ($\geq 1$ kHz) of the laser source is necessary in order to improve the photoelectron count and, at the same time, to minimize the space charge effects. Specifically, to achieve higher angular resolution and efficiency, the angle-resolved time-of-flight (ARTOF) analyzer has been developed[26,28,233]. As schematically shown in Fig. 6a, a typical ARTOF analyzer (e.g. ARTOF-10k from Scienta-Omicron[26]) consists of several cylindrical electrostatic lenses with some defined angular acceptance, which image the emitted electrons onto a delay-line detector positioned at the end of the lens. The kinetic energy and emission angle

(momentum) of a photoelectron are given by a combination of its flight time and its spatial striking position on the detector. This gives the possibility of collecting energy-momentum data for a complete area of the Brillouin zone ($E(k_x,k_y)$) rather than along a line $E(k_x)$, as with a traditional hemispherical analyzer[26]. Therefore, higher efficiency is achieved. We note that traditional ARTOF analyzer (e.g., ARTOF-10K) has an additional restriction of at most one electron detected per pulse due to the limitation of the delay-line detector. However, new types of delay line detectors have been recently developed[237] to overcome this limitation by allowing multiple electrons detected simultaneously per pulse. In practice, when a pulsed laser with repetition rate below 1 MHz is employed, the space charge effects, as discussed earlier, could be very pronounced even before the count rate of one electron per pulse is reached. Hence, for high energy-resolution measurements, minimizing the space charge effects is still the dominant consideration. To summarize, relatively high efficiency of ARTOF analyzer makes it an ideal detector for pulsed lasers. However, the inherent requirement of pulsed beam with appropriate repetition rate (< 3 MHz) restricts its application to other light sources, including synchrotron light sources, discharge gas lamps, and quasi-continuous lasers. In recent years, thanks to the rapid development of high repetition rate ($\geq$ 1 kHz) femtosecond and attosecond lasers[56,57], time-resolved ARPES has become increasingly popular with a variety of time-resolved ARPES systems involving different ranges of energy resolution (~ tens of to hundreds of meV), time resolution (~ hundreds of attosecond to hundreds of femtosecond), and pump fluence (~ tens of $\mu J/cm^2$ to several $mJ/cm^2$) have been developed. Note that there is a trade-off between time and energy resolution due to the Heisenberg uncertainty principle, e.g., a 150 femtosecond laser pulse would have an energy resolution > 12 meV.

Time-resolved ARPES is a direct tool to probe the dynamics of transient states and unoccupied states of the band structure, which is clearly illustrated in the study of grey arsenic[159]. Grey arsenic exhibits nontrivial Rashba-split Shockley states on the (111) surface, as shown in Fig. 6d. To pin down the nontrivial band topology of these Shockley states, time-resolved ARPES measurements have been performed on the (111) surface of grey arsenic. As illustrated in Fig. 6b, the ARPES data clearly reveal a pair of parabolic bands, which split along both $\overline{\Gamma}-\overline{M}$ and $\overline{\Gamma}-\overline{K}$ directions but are degenerate at the $\overline{\Gamma}$ point. In the $\overline{\Gamma}-\overline{K}$ direction, one band disperses into the conduction band, while the other turns back and merges into the valence band. Such a surface-to-bulk connection provides direct evidence for the nontrivial band topology of the Shockley states. Furthermore, by varying the pump-probe delay time, the dynamics of these unoccupied states is revealed in Fig. 6c. It is worth mentioning that besides the time-resolved ARPES, the unoccupied state can also be detected by two-photon photoemission (2PPE) spectroscopy[234–236,238–241], in which, a photon first promotes an electron from below the $E_F$ to an unoccupied intermediate state, and a second photon subsequently excites the hot electron above $E_{vac}$. Note that the 2PPE spectroscopy is usually performed with photon energies less than the work function, so that the intense one-photon photoemission (1PPE) signal can be avoided. Recently, J. A. Sobota et al.[149] reported the performance of both the 1PPE and 2PPE spectroscopies with single 6 eV (above the work function) femtosecond lasers by tuning the pulse intensity, as the 1PPE and 2PPE signal scale linearly and quadratically with pulse intensity, respectively.

More astoundingly, time-resolved ARPES can also be used to manipulate the electronic structure and potentially engineer new quantum states. Among the most intriguing achievements has been the observation of Floquet-Bloch states[123,127,242] in $Bi_2Se_3$. As known from Bloch's theorem, a spatially periodic potential in lattice results in the replication of band dispersion in momentum, namely, Bloch states. In analogy to Bloch states, a temporally periodic electromagnetic field (usually can be created by an intense pump laser pulse) leads to 'replicas' of the bands in energy, known as Floquet-Bloch states. The Floquet-Bloch states were recently demonstrated on the surface of a $Bi_2Se_3$ by using a pump pulse with energy below the bulk band gap[123,127]. As shown in Fig. 6f, without pumping, a single Dirac-cone surface state resides inside the bulk band gap. When pumping $Bi_2Se_3$ with $p$-polarized (configuration shown in Fig. 6e) laser pulses, the periodic duplication of these Dirac bands begins to appear along the energy axis, as shown in Figs. 6g-h. Interestingly, these dispersive band replicas cross with one another upon moving away from the $\bar{\Gamma}$ point, giving rise to new Dirac points along the $k_x$ direction (Fig. 6g). However, these new Dirac cones are gapped out along the $k_y$ direction (Fig. 6h,i), indicating the broken time-reversal symmetry induced by the $p$-polarized laser pulse. The above observations provide compelling evidence of the photon-dressed Floquet–Bloch states in solids and may pave the way for optical manipulation of new phases.

The capability of time-resolved ARPES is far beyond the above two examples, and it has been widely used in studying the momentum-resolved electronic dynamics of many types of materials, including high-temperature superconductors (e.g., cuprates[116,117,122,131–138] and iron-based superconductors[128,139–142,155]), density wave systems[114,115,118,145,146,156,157], topological insulators[119,121,123,127,147,148,150–154], graphene[124–126,130], and other strongly-correlated systems[143,144]. For example, besides the above discussed Floquet-Bloch states, dynamics of scattering between surface and bulk states has also been observed in topological insulator $Bi_2Te(Se)_3$[119,121,147]. Apart from transient dynamics, collective modes have also been observed in many materials, such as rare-earth tritellurides[115,146,156] and FeSe/STO films[155].

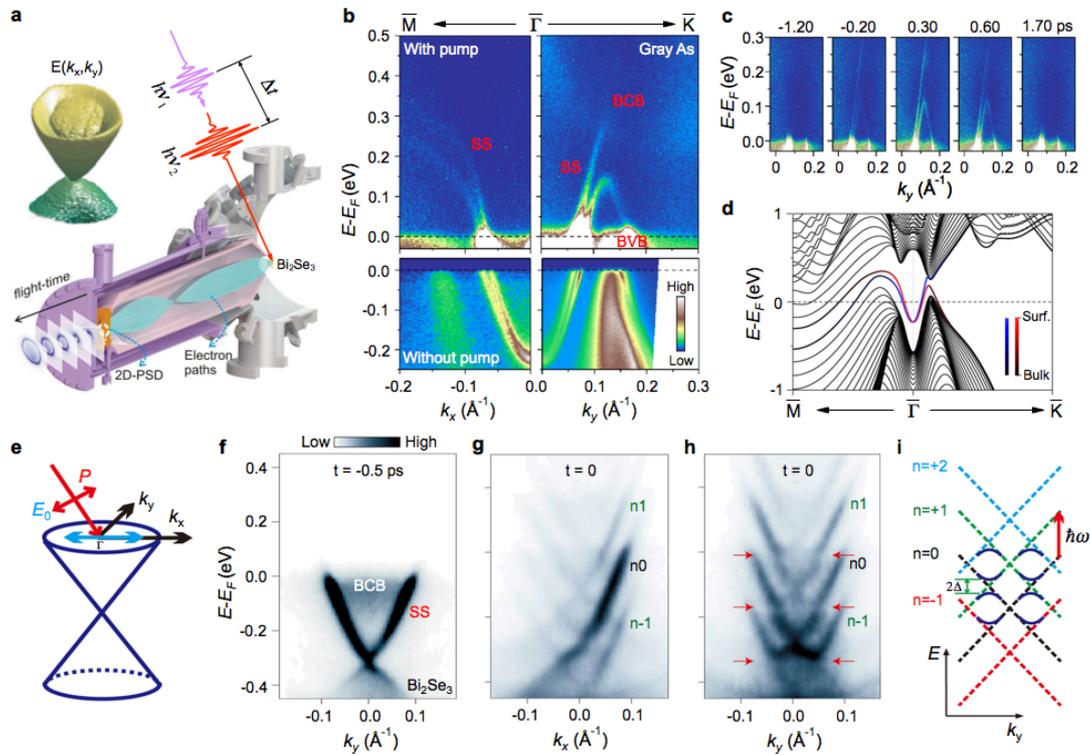

Figure 6 | Time-resolved ARPES studies of gray arsenic and $Bi_2Se_3$. **a** | Schematic of the time-resolved ARPES system with ARTOF spectrometer from Prof. Gedik's group at MIT. PSD: position-sensitive detector. Top left: a typical isointensity surface in ($E$, $k_x$, $k_y$) space from $Bi_2Se_3$. **b** | Measured band dispersions of gray arsenic below and above $E_F$ along $\overline{M}-\overline{\Gamma}-\overline{K}$ direction. Top: band dispersions above $E_F$ recorded by time-resolved ARPES measurements. Bottom: band dispersions below $E_F$ recorded by high-resolution laser ARPES measurements. **c** | Time-resolved ARPES intensity snapshots of gray arsenic along $\overline{\Gamma}-\overline{K}$ direction with various pump-probe delay times, showing the excitation or decay process of the electron states above $E_F$. **d** | Slab calculations of surface (indicated in blue and

red colors) and bulk bands (indicated in black color) along $\overline{M} - \overline{\Gamma} - \overline{K}$ direction. **e** | Sketch of the experimental geometry of time-resolved ARPES for the p-polarized pump laser. **f** | Time-resolved ARPES intensity plot of $Bi_2Se_3$ along the $k_y$ direction at t = -0.5 ps with p-polarized pump. **g** | Time-resolved ARPES intensity plot of $Bi_2Se_3$ along the $k_x$ direction at t = 0 with p-polarized pump. **h** | Same as **g**, but along the $k_y$ direction. Red arrows indicate the avoided crossing gaps. **i** | Sketch of the 'dressed' replica bands of different order as induced by the midinfrared excitation. Avoided crossing occurs along $k_y$ direction, leading to a band gap of $2\Delta$. Panel **a** is adapted with permission from ref.[233], American Physical Society. Panels **b-d** are adapted with permission from ref.[159], American Physical Society. Panels **e** and **i** are adapted with permission from ref.[123], AAAS. Panels **f-h** are adapted from ref.[127], Springer Nature Limited.

## Spatially resolved ARPES

Typically, the conventional synchrotron-based ARPES systems have a spatial resolution of ~ 100 μm, depending on the spot size of the light source (table 1), which implies that the measured sample should have a flat surface dimension ≥ 100 μm. However, many interesting materials or domains are below 100 μm in size, such as heterostructures, micro- and nano-scale materials. Even though the size of materials is above 100 μm, the flat regions of a cleaved surface may have dimension < 100 μm. In consideration of the above cases, spatially resolved ARPES with sub-micrometer or nanometer spatial resolution has been developed recently at several synchrotron light sources[31,96–100].

Spatially resolved ARPES may be viewed as a combination of ARPES and scanning photoemission microscopy. Depending on the beam spot size achieved, spatially resolved ARPES can be called 'micro-ARPES' or 'nano-ARPES' for the micrometer or nanometer scale beams, respectively. To focus the X-rays to a micro- or nano-size spot at the sample, advanced optics has been developed, such as Schwarzschild optics[96] and the Fresnel zone plate (FZP)[97]. As an example, Fig. 7a shows a schematic of the nano-ARPES end-station[99] at the MAESTRO beamline at the Advanced Light Source (ALS). The FZP system is used to focus the beam, together with an order selection aperture placed between the FZP and the sample to eliminate higher diffraction orders. Moreover, a high-precision sample stage is also needed to ensure precise nanometer scanning and positioning of the sample. The final spatial resolution is determined by the FZP resolution and the mechanical and thermal stability of the sample stage.

The development of spatially resolved ARPES makes it possible to map out the band structures of materials with micrometer or nanometer spatial resolution, while keeping the advantages of conventional ARPES, such as energy and momentum resolution. However, spatially resolved ARPES also has quite obvious shortcomings, including low count rates due to low focusing efficiency (~ 1%) of focusing optics, strong space charge effects due to nano-size beam spot, and limited photon-energy choices. Furthermore, it should be noted that there is a complementary technique, photoemission electron microscopy (PEEM)[243], which also allows spatially resolved ARPES measurements. However, the corresponding energy and momentum resolution (typically above 100 meV and 1°) of PEEM are usually worse than the ones of spatially resolved ARPES.

Spatially resolved ARPES have been successfully applied in the study of electronic structure of micro- or nano-scale materials or domains, including heterostructures[101–110], $Sb_2Te_3$ nanowires[111,112], and weak topological insulator $β$-$Bi_4I_4$[113]. Here we highlight the recent application of spatially resolved ARPES in the study of the weak topological insulator $β$-$Bi_4I_4$. As pointed out by L. Fu et al.[244,245], in three dimensions, a topological insulator can be classified as either 'strong' or 'weak' depending on the $Z_2$ topological invariants. The strong topological insulator, which manifests with gapless topological surface states at all the surfaces, has been experimentally confirmed in many materials, such as $Bi_{1-x}Sb$[161], $Bi_2Se(Te)_3$[91,92,162–164]. By contrast, the weak topological insulator state is very challenging to detect, because the corresponding gapless surface states emerge only on particular surfaces, which are typically undetectable in real 3D crystals. Very recently, using nano-ARPES, Ryo Noguchi et al.[113] provided a direct experimental evidence for the existence of weak topological insulator state in $β$-$Bi_4I_4$, by observing the topological Dirac surface states on the side (100) surface. As schematically shown in Fig. 7b, the weak topological insulator $β$-$Bi_4I_4$ exhibits no topological surface states on the (001) surface, but quasi-one-dimensional (1D) topological surface states on the side (100) surface. These topological surface states exhibit two Dirac cones at the $\bar{\Gamma}$ and $\bar{Z}$ points (Fig. 7c), which, in principle, can be directly detected by ARPES. However, as shown in Fig. 7d and e, $β$-$Bi_4I_4$ single crystals usually have very small size (~ 30 μm) along the stacking direction (c-axis), also, the cleaved (100) surface is actually composed of several small domains or stages (typically around 2 μm). Therefore, it is very difficult to study the surface states on the (100) surface by conventional ARPES, as the spot size (~ 100 μm) of incident beam is much larger than 2 μm. To examine the Dirac surface states exclusively, spatially resolved ARPES measurement with a beam spot less than 1 μm, is performed on the brightest region of (100) surface (white circle in Fig. 7e). As a result, the expected gapless Dirac dispersions are successfully resolved at both the $\bar{\Gamma}$ and $\bar{Z}$ points (Fig. 7f), demonstrating the experimental realization of the weak topological insulator state in $\beta$-$Bi_4I_4$.

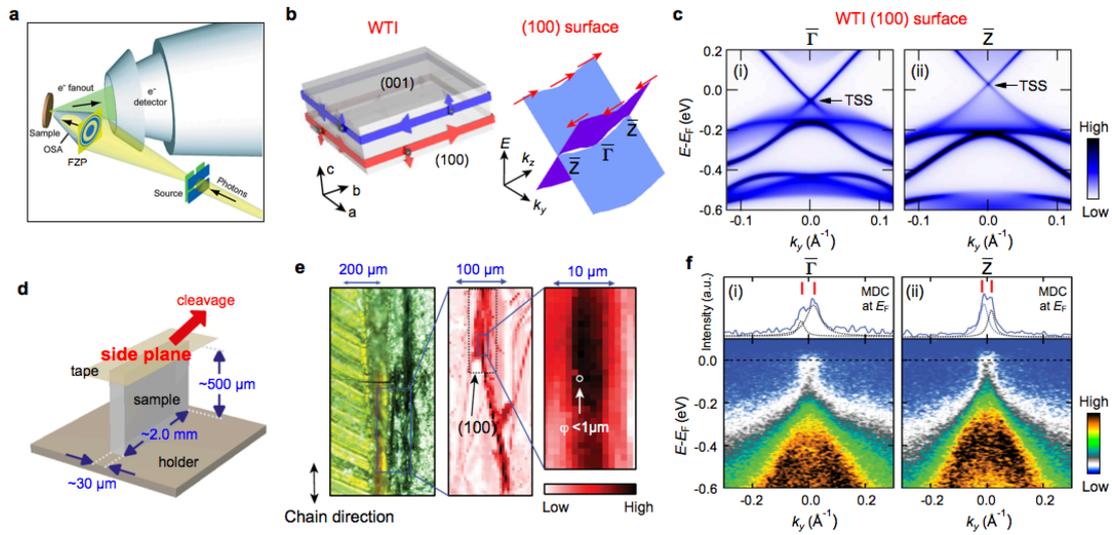

Figure 7 | Topological surface Dirac cone on the alternately doped graphene p-n junctions measured with spatially resolved ARPES. **a** | Schematic of the nano-ARPES system from Prof. Rotenberg's group at ALS. FZP: Fresnel zone plate, OSA: order selection aperture. **b** | Schematics of the topological surface states (TSSs) of weak topological insulator (WTI) $\beta$-$Bi_4I_4$ in real space and their band dispersions in k-space. The red arrows indicate the spin texture in k- space. **c** | Calculated Dirac band dispersions of TSSs at the $\bar{\Gamma}$ (c(i)) and $\bar{Z}$ (c(ii)) points of the side (100) surface. **d** | Schematic of the side (100) surface preparation for nano-ARPES measurements. **e** | Left, real-space image of the measured $\beta$-$Bi_4I_4$ sample taken by a optical microscope; centre, photoemission intensity map; and right, enlarged image. The white circle indicates the selected position for nano-ARPES measurements. **f** | ARPES intensity plots, as well as the momentum distribution curves (MDCs; blue curves) at the $E_F$, along the $k_y$ direction at the $\bar{\Gamma}$ (f(i)) and $\bar{Z}$ (f(ii)) points of the side (100) surface.. The black dotted curves are the Lorentzian fitting of the two peak structures (red bars) in the MDCs. Panel **a** is adapted

with permission from ref.[99], International Union of Crystallography. Panels **b-f** are adapted with permission from ref.[113], Springer Nature Limited.

## Conclusions and prospects:

With the unique capability of directly visualizing and discriminating the surface and bulk electronic states, modern ARPES has played a critical role in the study of topological materials. In this topical review, we summarize the fundamentals, advancements and challenges of different types of ARPES techniques, including synchrotron-based surface-sensitive VUV ARPES and bulk-sensitive soft X-ray ARPES, laser-based high-resolution ARPES and time-resolved ARPES, spin-resolved ARPES, and spatially resolved ARPES. Meanwhile, we also highlight several scientific applications on topological materials to better demonstrate the capabilities of these different types of ARPES techniques.

In recent years, continuous efforts have been made to improve the capabilities of ARPES. To further reduce the sample temperature to below 1 K, new cryostats have been developed, such as He-3 cryostats[30]. To increase the performance and data acquisition efficiency, various types of electron analyzers are developed, such as DA30-L and ARTOF analyzers[26], which allow the detection of the 2D in-plane electronic structure, $E(k_x,k_y)$, at a fixed sample geometry. For light sources, many advanced synchrotron-based beamlines are available now, including the "Dreamline" at Shanghai Synchrotron Radiation Facility (SSRF) with high energy resolution (~ 25meV @ 1 keV) and wide photon energy range (20-2000 eV), the I05 beamline[35] at Diamond and the beamline 7.0.2[99] of the Advanced Light Source (ALS) at the Lawrence Berkeley National Laboratory with nano-scale beam spot sizes. We also witness the emergence of a newly-developed light source: X-ray free-electron lasers[246–252]. The X-ray free-electron laser might be a promising light source for time-resolved ARPES since it features broad tunable and femtosecond pulsed beam with high photon flux. In particular, the broad tunable light makes it possible to perform systematic time-resolved ARPES measurements over the full 3D Brillouin zone, which is crucial for 3D materials such as topological semimetals. Currently，the low repetition rate (about tens to hundreds of Hz) of these free-electron lasers is a major restriction for the applications of time-resolved ARPES, because the pulse intensities must also be kept low to minimize space charge effects. Consequently, the photoelectron count is not sufficient for efficient high-statistics acquisition of time-resolved spectra.

Looking forward, there is still plenty of room to improve the capabilities of ARPES. High momentum and energy-resolution laser ARPES with He-3 cryostats is needed to probe the superconducting gap of low-$T_c$ superconductors, which is already under development in several laboratories. Despite great interest, spin-resolved soft X-ray ARPES is still not realized due to the poor efficiency of spin detectors and low cross sections of soft X-ray photoemission. The development of high-efficiency spin detectors and ultra-bright synchrotron light will make it possible to perform spin-resolved ARPES measurements in the soft X-ray regime. Similarly, with continuing improvements of focusing optics, we expect that the efficiency of spatially resolved ARPES will increase. For time-resolved ARPES, it is still very important to develop high-repetition rate and stable lab-based lasers with tunable photon energy and high photon flux. At the same time, developing strong tunable pump pulses from the ultraviolet down to the terahertz range is important in the study of coherent excitations of low energy modes such as lattice vibrations, and various quasiparticle excitations in nontrivial band structures. Indeed, the first time-resolved ARPES system with terahertz pump laser has been developed very recently[154], and successfully applied to the study of Dirac fermions in $Bi_2Te_3$. Furthermore, the combination of high-efficiency spin detectors with time-resolved ARPES systems would provide new opportunities to acquire simultaneous time- and spin-resolved ARPES data. In particular, we note that a high efficiency ARTOF-VLEED spin spectrometer, which combines the VLEED spin detector with the ARTOF analyzer, has been developed recently[76,82]. This high-efficiency spectrometer may offer a promising platform for simultaneous spin-, time-, and angle-resolved photoemission with pulsed lasers. Lastly, the interconnection of ARPES with sample synthesis systems (e.g., MBE, PLD) and other complementary *in-situ* characterization instruments (e.g., scanning tunneling microscope, transport, and optical measurements systems) would provide an efficient platform to symmetrically design and synthesize novel materials, as well as to investigate their physical properties with different experimental methods.

ARPES has played an irreplaceable role in discovering and understanding the unique physical properties of many topological materials, including topological insulators[91,92,161–166], topological Dirac[172–176] and Weyl semimetals[65,66,69,177–184], and unconventional fermions[67,68,185–193]. Beyond these, there are still several prominent physics problems waiting to be explored by ARPES. For example, the spin texture of bulk Weyl cones has not been investigated due to a lack of spin-resolved soft X-ray ARPES system. What's more, many predicted topological phases still await strong ARPES evidence such as magnetic Weyl semimetals[177,178]. Very recently, the discovery of superconductivity in twisted bilayer graphene[253,254] has attracted extensive attention, the interesting electronic structure might be resolved by spatially resolved ARPES. Hence, we strongly believe that the ARPES technique will continue to play a leading role in the research of novel topological materials in the future.

## Acknowledgments

The authors are indebted to A. Zong, B. Fichera, C. Belvin, N. Koirala, and Y. Su for useful discussions and feedback in the preparation of this manuscript. This work was supported by the Ministry of Science and Technology of China (2016YFA0401000, 2016YFA0300600, and 2015CB921300), the Chinese Academy of Sciences (XDB28000000, XDB07000000, and QYZDB-SSW-SLH043), the National Natural Science Foundation of China (11622435 and U1832202), and the Beijing Municipal Science and Technology Commission (No. Z171100002017018). B.Q.L. acknowledge support from the National Science Foundation under Grant No. NSF DMR-1809815 (data analysis), and the Gordon and Betty Moore Foundation's EPiQS Initiative grant GBMF4540 (manuscript writing).

## Author contributions

All authors made contributions to the discussion of contents and researched data for the article. All authors participated in writing of the manuscript, with B.Q.L. contributing the most.

## Competing interests

The authors declare no competing interests.

**Direct experimental evidence from spin-resolved ARPES that topological insulators possess spin-momentum locking feature.**

**Experimental realization of the first 3D topological insulator $Bi_{0.9}Sb_{0.1}$.**

**High-resolution laser-ARPES measurements provide direct evidence for the existence of topological superconductivity on the (001) surface of an iron-based superconductor.**